\documentstyle[referee,epsfig]{mn}
\def\be{\begin{equation}}
\def\ee{\end{equation}}
\def\bea{\begin{eqnarray}}
\def\eea{\end{eqnarray}}
\def\etal{{et al.}\thinspace}

\begin{document}

\title
[Cooling flows and the entropy of the intragroup medium
]
{Cooling flows and the entropy of the intragroup medium}

\author[Biman B. Nath]
{Biman B. Nath\\
Raman Research Institute, Bangalore 560080, India\\
(biman@rri.res.in)
}
\maketitle

\begin{abstract}
{
We study steady, homogeneous and subsonic cooling flows 
in poor clusters of galaxies
in light of the recent proposal that radiative cooling of the intracluster
gas can explain the observations of the `entropy floor' and other related
X-ray observations. We study a family of cooling flow solutions
parameterized by the mass flux rate, and then determine the resulting entropy and
the X-ray luminosity. We find that cooling flows with mass flux rates
in excess of $1000$ M$_{\odot}$ per yr are required to explain the 
observations of entropy and X-ray luminosity. In view of the observed lack of
such large flows in rich clusters, our calculations
suggest that heating sources are needed, in addition to the effect of
radiative cooling, to explain the observations.
}
\end{abstract}


\section{Introduction}
Studies of the structure of dark matter in collapsed objects of different
masses show a remarkable universality. Numerical simulations have shown
that the dark matter density profile has a universal shape (Navarro, Frenk
\& White 1995). This result provides a framework for the study of the
relation between the baryonic and dark matter content of halos. Clusters
of different masses, with their large baryonic gas content, are excellent
probes for such studies.
If the physical properties of the baryonic gas contained
in these self-similar potential wells were completely governed by gravity,
then the observed properties of the gas would also be scale-free (Kaiser
1986). For
example, relation between the gas temperature and the cluster mass
or the X-ray luminosity and the temperature, would be independent of the
total mass ($T \propto M^{2/3}$ and $L \propto T^2$, respectively).

Observations, however, show that in reality such relation deviate from
being self-similar, pointing out the importance of non-gravitational processes
determining the properties of the baryonic gas. The observed relation
between luminosity and temperature steepens from the self-similar
expectation of $L \propto T^2$ for low
temperature clusters (e.g., Markevitch 1998; Allen \& Fabian 1998). 
Also the mass-temperature
relation shows deviation from self-similarity (e.g., McCarthey \etal 2002).

It has also been observed that the surface brightness profiles of poor
groups (with lower temperatures than richer clusters) are relatively shallow
(Ponman \etal 1999), showing that the underlying gas density profile is also
shallow. All these observations clearly point towards some physical process
which raises the entropy of the gas. Indeed, direct determination of the
gas entropy (one defines an `entropy' $S=T/n_e^{2/3}$ which is the entropy
of an ideal gas, stripped of the logarithm and the constant factors)
shows the existence of an `entropy floor', a minimum level of entropy.
Lower temperature clusters, therefore, have relatively high entropy gas,
compared to the expectations from self-similar models. It has been
estimated that this `entropy floor' corresponds to $\sim 100$ keV cm$^{-2}$
(Ponman \etal 1999).

One way to achieve this high level of entropy is to pre-heat the gas
(Kaiser 1991), and previous authors have estimated that extra non-gravitational
heating of order $\sim 1$ keV per (gas) particle should suffice to explain
the observations. The natural choices for the heating agents are supernovae
driven winds (e.g., Ponman \etal 1999) and active galactic nuclei (e.g, 
Yamada \& Fujita 2001;
Nath \& Roychowdhury 2002), although uncertainties prevail in the estimates
of their effects on the intracluster medium.

The other alternative is to siphon off the lowest entropy part of the gas
through cooling, from the diffuse intracluster medium (ICM) to some other
phase, so that it does not enter the calculations of the X-ray luminosity
or X-ray weighted temperature of the gas. Voit \& Bryan (2002) recently
argued on the basis of the cooling function for a metallicity of
$Z=0.3 \, Z_{\odot}$ that radiative cooling can be an efficient process (given
a Hubble time to operate on) in introducing an entropy floor of the
appropriate level. Voit \etal (2002) have studied the effect of different
kinds of entropy modifications on the properties of the cluster gas.
As an example of such an entropy modification, they considered the effect of
radiative cooling, assuming that it would introduce an entropy threshold
corresponding to the entropy at the cooling radius (where cooling time
equals the age of the system). They
found that profiles with such an entropy threshold can
adequately explain the observations.
Wu \& Xue (2002b) also considered the effect of radiative cooling by removing
the parcel of gas whose cooling time scale is less than the age of the
cluster, which would cause the outer layers of gas to flow inside to
again assume a hydrostatic profile. Their calculation, however, assumes
that the entropy of the outer gas remains unchanged as it flows inward.
Their predicted fraction of cool gas, however, exceeds the observed stellar
mass fractions estimated by Roussel, Sadat \& Blanchard (2000).

It is clear that radiative cooling would set the intracluster gas 
in motion towards the center (e.g., Silk 1976, Fabian 1994, and more recently
B\"ohringer \etal 2002, and references therein).  
Since the X-ray properties of the gas mostly
depend on the density and temperature profile of the gas, it is important
to determine these profiles as accurately as possible, taking into account
the changes brought about by such a flow.  

%Assuming hydrostatic
%equilibrium profiles at the beginning and after a Hubble time can only
%be an approximation. Also, it is important to take into account the change
%in entropy of the inward flowing gas.

Several authors have performed numerical simulations of
cooling flows in clusters of different masses in light of the deviant
X-ray properties of poor cluster, for example, Knight \& Ponman (1997),
Dav\'e \etal (2002) and Muanwong \etal
(2002).  Knight \& Ponman (1997) found that
the density and temperature profiles caused by cooling flows were not
sufficient to explain the X-ray properties, although they had used
a $\Omega_0=1$ cosmology with a dark matter profile from Bertschinger (1985),
different from the universal profile used by more recent authors.
Also, they did not address the problem of the entropy `floor' as it
was not known then. More recently Muanwong \etal (2002) found that cooling
flows which come about as a result of radiative cooling can reproduce all
observations, whereas Dav\'e \etal (2002) find some discrepancy in their
results compared to the observations.

In this paper, we study simple models of cooling flows for
the effect they have on the X-ray
properties of clusters, and in setting up the 'entropy floor'. We simplify
the cooling flow equations motivated by the recent X-ray observations
(e.g., B\"ohringer \etal 2002). We do not have any heating
sources in the flow equations, so that we can study the effect of radiative
cooling alone. 

We begin by setting up the background dark matter potential and the default
profiles of the gas in the next section. We also introduce the cooling flow
equations later in the section. We present the result of our calculations
in \S 3 and discuss the implications, along with the uncertainties and 
limitations of our calculations in \S 4.

We assume throughout the paper that $\Omega_\Lambda=0.7$, $\Omega_0=0.3$
and $h=0.65$.

\section{Density profile}
We first consider the density profile of the dark matter which provides
the potential for the gas particles.

\subsection{Dark matter density profile}
We assume that the gas mass is negligible compared to that of the total
dark matter mass in the clusters, and that the dark matter density profile
is given by the `universal' profile, expressed in terms of a
characteristic radius $r_s$ (e.g., in Komatsu \& Seljak 2002),
\be
\rho_{dm}(r)=\rho_s y_{dm}(r/r_s) \,,
\ee
where $\rho_s$ is a normalizing density parameter and $y_{dm}(x)$ is given by
\be
y_{dm}(x)={1 \over x^{\alpha} (1+x)^{3-\alpha}} \,.
\ee
Here the parameter $\alpha$ characterizes the shape of the profile.
The total dark
matter mass within a radius $r$ is
\be
M(\le r)=4 \pi \rho_s r_s^3 m(r/r_s) \,,
\label{eq:totm}
\ee
where,
\be
m(x)=\int_0^x du \, u^2 \, y_{dm}(u)=\ln (1+x)-{x \over 1+x} \,;
\ee
Here, the last equality is valid for $\alpha=1$ which is the much used NFW
profile (Navarro \etal 1996, 1997). 
Recently, Moore \etal (1998) and Jing
\& Suto (2000) thought that $\alpha=1.5$ better described their simulation
results.  For $\alpha=1.5$, $m(x)=2 \ln (\sqrt{x}+\sqrt{1+x})-2 \sqrt{{x
\over 1+x}}$ (Suto \etal 1998). 
The characteristic radius $r_s$ is related
to the virial radius $r_{vir}$ by the `concentration parameter' ($c$), as
\be
c \equiv {r_{vir} \over r_{s}} \,.
\ee
The total mass of the cluster is assumed to be the mass inside its
virial radius.
The virial radius is calculated in the spherical collapse model to be,
\be
r_{vir}=\Bigl [ {M_{vir} \over (4 \pi /3 )
\Delta_c (z) \rho_c(z) } \Bigr ]^{1/3}
=\Bigl[ { M_{vir} \over 4 \pi c^3 \rho_s m(c) } \Bigr ]^{1/3} \,,
\ee
where the second equality comes from evaluating equation (\ref{eq:totm})
at the virial radius. Here $\Delta_c(z)$ is the present day
 spherical overdensity of
the virialized halo within $r_{vir}$ at $z$ in the units of the critical
density of the universe $\rho_c(z)$. Following Komatsu \& Seljak (2002),
we assume a value $\Delta_c (z=0)=100$ for a cosmological model with
$\Omega_m=0.3$ and $\Omega_\Lambda=0.7$.

We follow Seljak (2000) in adopting  the
approximation for $c$ as a function of the
cluster mass,
\be
c=6 \Bigl ( {M_{vir} \over 10^{14} h^{-1} \, M_{\odot}} \Bigr )^{-1/5} \,.
\ee
The above set of equations specify the dark matter density profile given the
mass of the cluster. We now turn our attention the
profile assumed by gas in these
clusters.

To compare our results with observations, which usually use the radius
$r_{200}$ where the overdensity is $200$, we compute in each case this
radius, and present our results in its terms.

\subsection{Gas density profile}
Our aim is to determine the changes
in the properties of the cluster which come about as a result of radiative
cooling. For this we need to assume a `default' profile of the gas and
then study the changes wrought upon the profile by these processes.

We assume that the default profile is that of gas in hydrostatic equilibrium
with the background dark matter potential. This does not specify the gas
density profile as more constraints are needed. One choice is that
the gas obeys a polytropic equation of state. This has been shown to be
provide an accurate description of the gas density and temperature profile
for rich clusters. 
Recent analyses find that gas profiles in hydrostatic equilibrium with a
polytropic index $\gamma \sim 1\hbox{--}1.2$, where pressure of the gas is
related to the density as $p \propto \rho ^{\gamma}$ describe the
observations well (Markevitch \etal 1998).
Since the rich clusters are thought to have been affected
minimally by non-gravitational processes, these profiles provide a natural
choice for the default profile. 
We note that this was also the approach of Wu, Fabian \& Nulsen (2000).

Recently, Komatsu \& Seljak
(2002) have determined the solutions to the hydrostatic equilibrium
equation, in the case of a polytropic gas, with the constraint that the gas
profile is proportional to that of the dark matter at large radii. This
physically motivated constraint limits the range of the polytropic index
$\gamma$, and they find that their solutions yield values of $\gamma$ within
the range between $1.1$ and $1.2$ for clusters with masses between
$10^{13} \hbox{--} 10^{15}$ M$_{\odot}$, as found in observations. 
They also provide useful
analytical fits for their solutions.

We adopt these profiles obtained by Komatsu \& Seljak (2002) as one 
set of our default
profiles (set A). We normalize these profiles by assuming that the 
ratio of the gas
mass inside the virial radius to the total virial mass is constant for all
clusters, and equals $\Omega_b /\Omega_m$. We adopt a value of $\Omega_b$
as constrained by primordial nucleosynthesis, as $\Omega_b \sim 0.02 h^{-2}$
(Burles \& Tytler 1999).

We also use another set of default profiles (set B), in which the gas density
is given by $\rho_g (r)=(\Omega_b / \Omega_m) \rho_d (r)$, being proportional
to the background dark matter density. This has been choice of several recent
authors, e.g., Bryan (2000), Voit \etal (2002) and others. 
The temperature profile is then given by the equation of hydrostatic equilibrium,
\be
{dp \over dr} =- \rho_g(r) { GM(r) \over r^2} \,.
\ee
where $M(r)$ is the total mass inside radius $r$.
We calculate the temperature profile of the gas with the 
boundary condition that the pressure is zero at infinity
(as in Wu \& Xue (2000a)). The resulting profiles are different
from that of Bryan (2000) only near the virial radius and does not affect
our calculations of the X-ray properties, since we compute these properties
only to an extent $r_{200}$, which is somewhat smaller than $r_{vir}$
for our choice of cosmology.

As we will find in the next section, the default temperature profiles for
set A and B differ substantially. The temperature profile for set A are
flat in the core region, whereas those for set B decline towards the centre.
The recent observations with the help of {\it CHANDRA} for the X-ray gas
fraction in clusters favour the default profiles for set A, with gas
fraction increasing with cluster mass (Allen, Schmidt \& Fabian 2002). 
In contrast, the gas fraction for set B is
constant by construction. Recent {\it CHANDRA}  
observations of temperature
profiles of some relaxed clusters do show declining temperature in the very central
region though, as in set B profiles (e.g, Schmidt, Allen \& Fabian 2001).

\subsection{Effect of cooling}
If we define a cooling radius, $r_{cool}$ as where the cooling time of the
gas equals the age of the cluster, then the gas inside this cooling radius
would cool rapidly.
It is widely
believed that in this case a cooling flow would ensue (Fabian 1994). 
Although it was generally believed in the last decade that cooling flows 
would have several phases, with gas being deposited at different radii,
recent observations find that such multiphase gas occurs only at the
very central region and for most of the region of the flow, one can
assume it to be homogeneous (e.g., B\"ohringer \etal 2002, and reference
therein). Also, recent observations rule out flows with very large
mass flux and gas temperatures dropping to very low temperatures, unless
at the very central regions, pointing towards the subsonic nature of the
flows.

Motivated by these points, we simplify the fluid equations by assuming a
steady, homogeneous and subsonic flow, 
in which the flow is dictated by the following
equations:
\bea
\dot{M}&=&4\pi r^2 \rho \, u \nonumber\\
{dp \over dr}&=&-\rho {d\phi \over dr} \nonumber\\
\rho \, u\,{d \over dr}(H+\phi)&=&n_e^2 \Lambda_N (T) \,,
\eea
where $H={5 \over 2}{k_B T \over \mu m_p}$ is the specific enthalpy,
$n_e$ is the electron density, $\Lambda _N(T)$ denotes the
normalized cooling function in the units of erg cm$^3$ s$^{-1}$, and $\phi$ is the
gravitational potential. The cooling function is normalized in a way
so that the total rate of cooling is given by $n_e n_i \Lambda _N(T)$ where
$n_i$ is the ion density (here we take $n_e=n_i$).

We assume a metallicity of $Z/Z_{\odot}=0.3$ for the ICM that is observed
in rich clusters and which has been used by all previous authors. We note
here that there is a large uncertainty in the abundance measurements for the
gas in poor clusters (Davis \etal 1999; Buote 2000). 
We use a fit to the normalized cooling
function for this metallicity as calculated by Sutherland \& Dopita (1993)
as described in Appendix A.

If we scale the density and temperature to their
values at $r_{cool}$, defining $T'=T/T_c$ (where $T_c$ is the temperature
at $r_{cool}$), $n'=n/n_{cool}$ and $x=r/r_{cool}$, also $\beta=r_{cool}/r_s$,
then the above
equations can be cast into a set of two equations with dimensionless
variables,
\bea
 {dT' \over dx}&=&n'^2 \, x^2 { t_{fo} \over  t_{co}} \,
{\Lambda(T) \over \Lambda(T_c)} - A  {(\ln (1+ x \beta) -{x \beta 
\over (1+x \beta)^2} ) \over (x \beta)^2}
\nonumber\\
{d n' \over dx}&=&-{A n'\over  T'} {(\ln (1+x \beta)-{x \beta \over 
(1+x \beta)^2} )\over (x \beta)^2} -{n' \over T'}
{dT' \over dx}\,,
\label{eq:coolfl}
\eea
where we have assumed $\alpha=1$ for the background dark matter profile, 
and where
\be
A=4 \pi G \rho_s r_s r_{cool} \mu m_p/k_B T_{c}
\ee
is a dimensionless quantity and is a measure of the ratio between the
potential and thermal energies at $r_{cool}$. Here $\mu$ is the mean
molecular weight. If $A$ is small, as is the
case for rich clusters, then gravity is negligible and one has a constant
pressure solution. In the last equation, we have defined the flow time at
$r_{cool}$ as,
\be
t_{fo}=r_{cool}/u_{cool}=4 \pi \rho_{cool}\, r_{cool}^3/\dot{M} \,,
\ee
and the cooling time at $r_{cool}$ to be,
\be
t_{co}={(5/2) n_{cool} k_B T_{c} \over n_{e, cool}^2 \Lambda (T_c)} \,.
\ee

\subsection{Family of solutions}
It is then possible to have a family of solutions for this equations for
different boundary conditions. 
%Our object, however, is to compare cooling
%flow solutions across the spectrum of cluster masses. For the sake of
%comparison, let us adopt a boundary condition which we will apply for all
%clusters. 
In the spirit of Nulsen \etal (1982; Appendix A), we will refer to 
the solution for which $T' \rightarrow 0$ as $x \rightarrow 0$
(meaning $T \rightarrow 0$ as $r \rightarrow 0$) as the `cooling 
eigensolution'. This boundary condition will fix the value of the ratio
${t_{fo} \over t_{co}}$, which will in turn fix the value of the mass flux.
The value of this ratio for the eigensolutions are of order unity as
expected (Nulsen \etal 1982).

Larger values of ${t_{fo} \over  t_{co}}$ 
than these particular values would then
yield solutions for which temperatures plunge to very low values at $r \le
r_{cool} \gg 0$. These profiles would give rise to X-ray holes in clusters
of galaxy, and we will not study them here.
Smaller values of the ratio lead to
solutions for which temperatures are larger than those for the
eigensolutions. 

At the other extreme, there are solutions to the cooling flow equations
with very small values of $(t_{fo} \over t_{co})$ corresponding to 
flows with very large mass flux rate. We find that the profiles for
such small values differ little if one decreases this ratio below
values of order $0.01$. One can actually get a formal limit to the
profiles by putting this ratio equal to zero in the cooling flow equations,
and numerically obtained profiles tend toward these limiting profiles
very rapidly below this value of $({t_{fo} \over t_{co}})$. For these
solutions, gas temperature is determined mainly by the gravitational
potential, as in this case $T(r) \propto H(r)= \phi(r)$
 (see also Fabian \etal 1984), especially for poor clusters
where gravity is important (where $A$ is large).
Such cases correspond to flows with mass flux rates that are $\sim 100$
times larger than the rates for eigensolutions.
We will refer to these set of limiting solutions as flows with
the highest entropy and very large mass flux rates. 

These two limits of the family of solutions parameterized by the
ratio $({t_{fo} \over t_{co}})$, or, equivalently, the mass flux rate,
constrain the physical solutions of the cooling flow equations relevant
for our purpose.

Gas beyond the cooling radius, $r_{cool}$ will remain at hydrostatic
equilibrium. For gas beyond this radius, we adopt the profiles as
discussed in the preceding section, which allow us to  calculate the extent
$r_{cool}$ and then $\rho_{cool}$ and $T_{c}$. We then use  these values in
equation (\ref{eq:coolfl}) to determine the cooling eigensolution for $r <
r_{cool}$.

We find that although the cooling eigensolutions are easily determined
for set A profiles, the transition from the case of $T=0$ at $r \gg 0$
to $T=0$ at $r=0$ happens over an extremely small range of values of
the ratio ${t_{fo} \over t_{co}}$, making it difficult to evaluate the
exact eigensolution. We have therefore used the lowest entropy profile,
with the lowest temperature possible at the central region, as the
eigensolutions for set B default profiles.

We perform two sets of calculations with the age of the clusters being
$10^{10}$ and $1.5 \times 10^{10}$ yr for the two sets. We have done
the calculations for cluster masses for which $r_{cool}/r_{vir} \ge 0.01$
where the cooling flows could have some tangible effects on the observable
properties of the gas. We have checked that
the effect of cooling flows on the total X-ray luminosity
is less than $\sim 20 \%$ for clusters with larger masses.

\section{Results}
\subsection{Gas density, temperature and entropy profiles}
We have computed the profiles for these two extreme cases for default
profiles in set A and B. An example of the gas density and temperature
profiles for a cluster mass of $3.5 \times 10^{13}$ M$_{\odot}$ is shown
in Figure 1, for a cooling time of $10^{10}$ yr.  
The top panels in Figure 1 show the density profiles, where the dotted
lines show the default profile, the solid slines show the lowest entropy
and the dashed lines show the highest entropy cases. The bottom panels
show the temperature profiles for the corresponding cases.

\begin{figure}
\centerline{
%{\vskip-4mm}
\epsfxsize=1.0\textwidth
\epsfbox{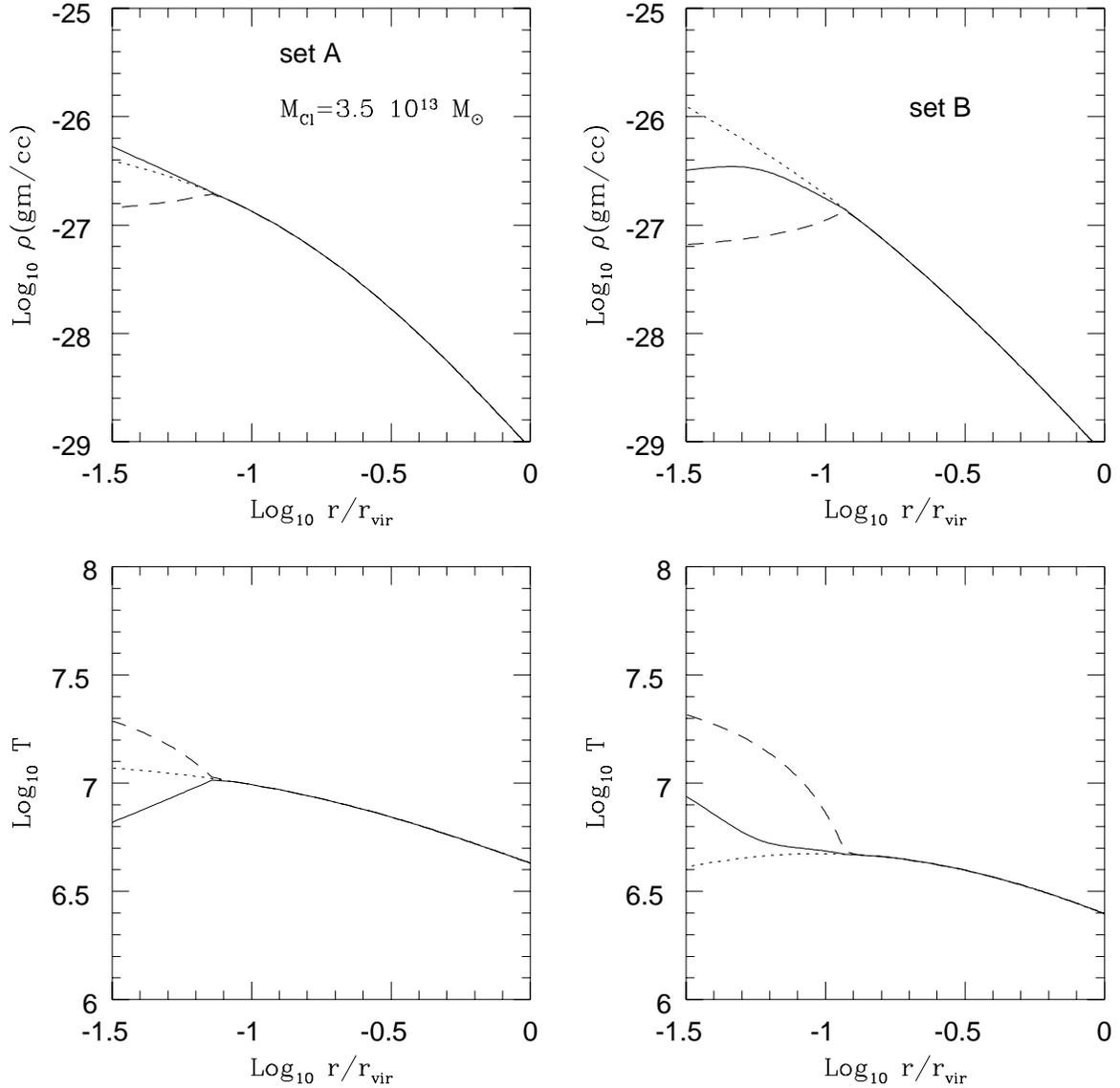}
}
{\vskip-3mm}
\caption{
Gas density (top panels) and temperature (bottom panels) are plotted
against $r/r_{vir}$
for cooling flows for set A (left panels) and set B (right panels)
default profiles, for a cluster mass of $3.5 \times 10^{13}$ M$_{\odot}$
and a cooling time of $10^{10}$ yr. 
Dotted lines in all panels show the default profile,
solid lines show the profiles for the `cooling eigensolutions' and the dashed
lines show the profiles with the highest mass flux rate.)
}
\end{figure}

An important difference between the set A and B cases is
that the default gas temperature at a given radius
in set B (in which the gas density
is proportional to the dark matter profile) is smaller than in set A. This
results in making the cooling radius larger for set B than in set A
(as the cooling function in the relevant temperature range 
increases with decreasing temperature).
This makes cooling flows somewhat more extensive for set B profiles.
Although this is true for clusters of a given mass, the emission weighted
temperature for set B profiles are smaller than those for set A profiles.
Since the emission weighted temperature is the observed parameter, and not
the mass, we find that for a given cluster temperature, set B profiles have
cooling flows with smaller extent.
We show the ratio $r_{cool}/r_{vir}$ as a function of the emission
weighted temperature (see below) in Figure 2 for the two sets of default
profiles and two choices of the cooling time.
Curves in Figure 2 also show that the cooling flows are more
extensive for lower temperature clusters as expected.

\begin{figure}
\centerline{
%{\vskip-4mm}
\epsfxsize=1.0\textwidth
\epsfbox{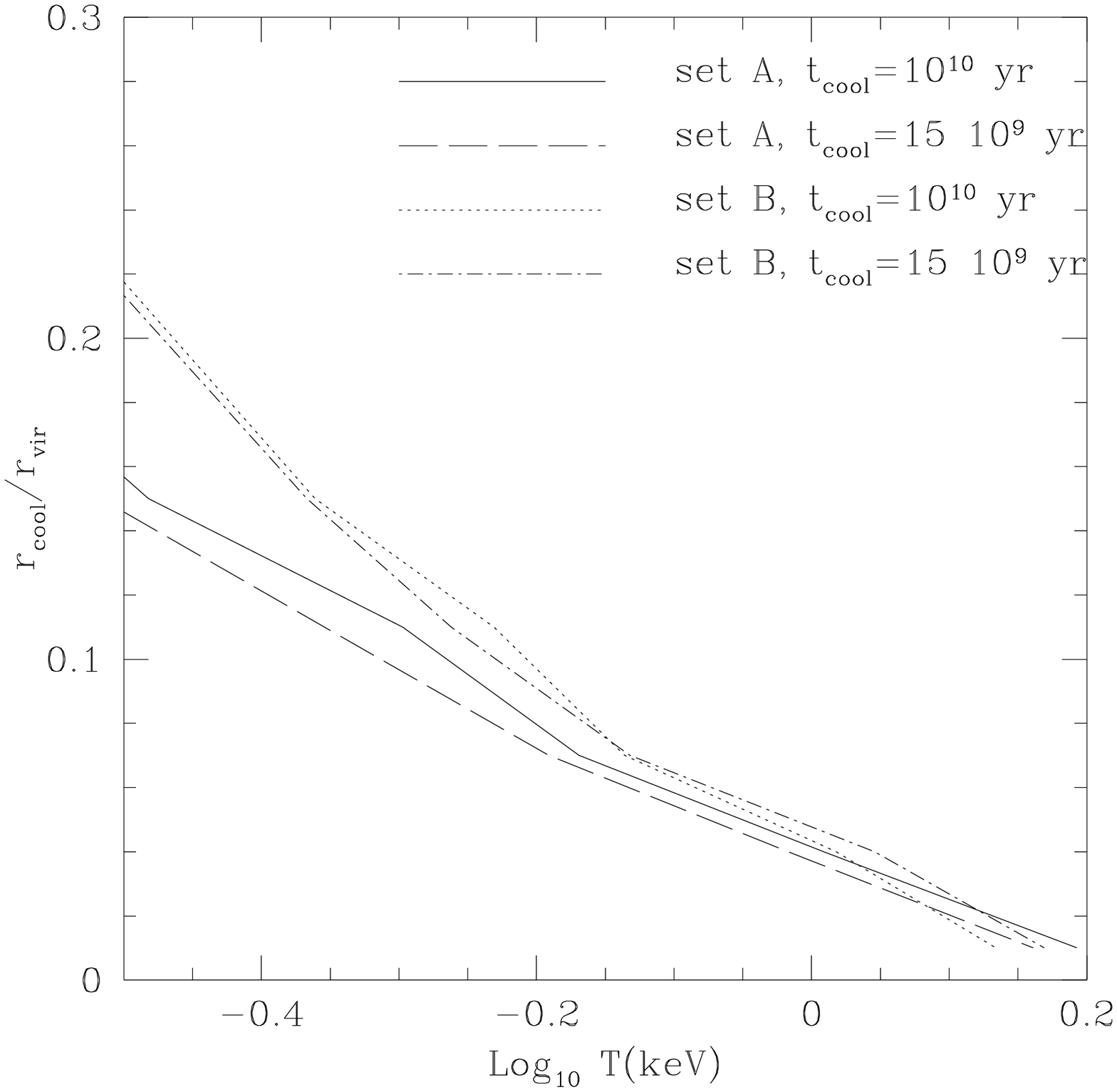}
}
{\vskip-3mm}
\caption{
The ratio of the cooling radius to the virial radius corresponding
to the two sets of cooling times ($10^{10}$ and $1.5 \times 10^{10}$ yr)
as a function of (emission weighted) cluster temperature.
}
\end{figure}

The corresponding entropy profiles are shown in Figure 3, which
shows more differences between the cases set A and B. Firstly, set A
profiles (with polytropic equation of state) have higher entropy to
begin with (upper dotted line). This is expected from other results
obtained by Komatsu \& Seljak (2000) regarding the shape of the profile.
They showed that their profiles assuming polytropic equation of state
could already explain the observed relation between the parameter $\beta$
(which defines the density profile). Observations show that $\beta$ increases
with cluster mass, showing that poor clusters have shallower profiles, which
is an indication of higher entropy. It is therefore expected that these
default profiles would have higher entropy than the other set of default
profiles. 

Secondly, cooling eigensolutions with the lowest entropy
(upper solid line) can decrease the entropy of the set A default
profiles. The default
profile for set B (lower dotted line), however, has very low entropy
to begin with, and
it is difficult to find solutions of the cooling flow with entropy lower
than this. The lowest entropy flows (lower solid line) in fact has larger
entropy than the default profile in this case.

\begin{figure}
\centerline{
%{\vskip-4mm}
\epsfxsize=1.0\textwidth
\epsfbox{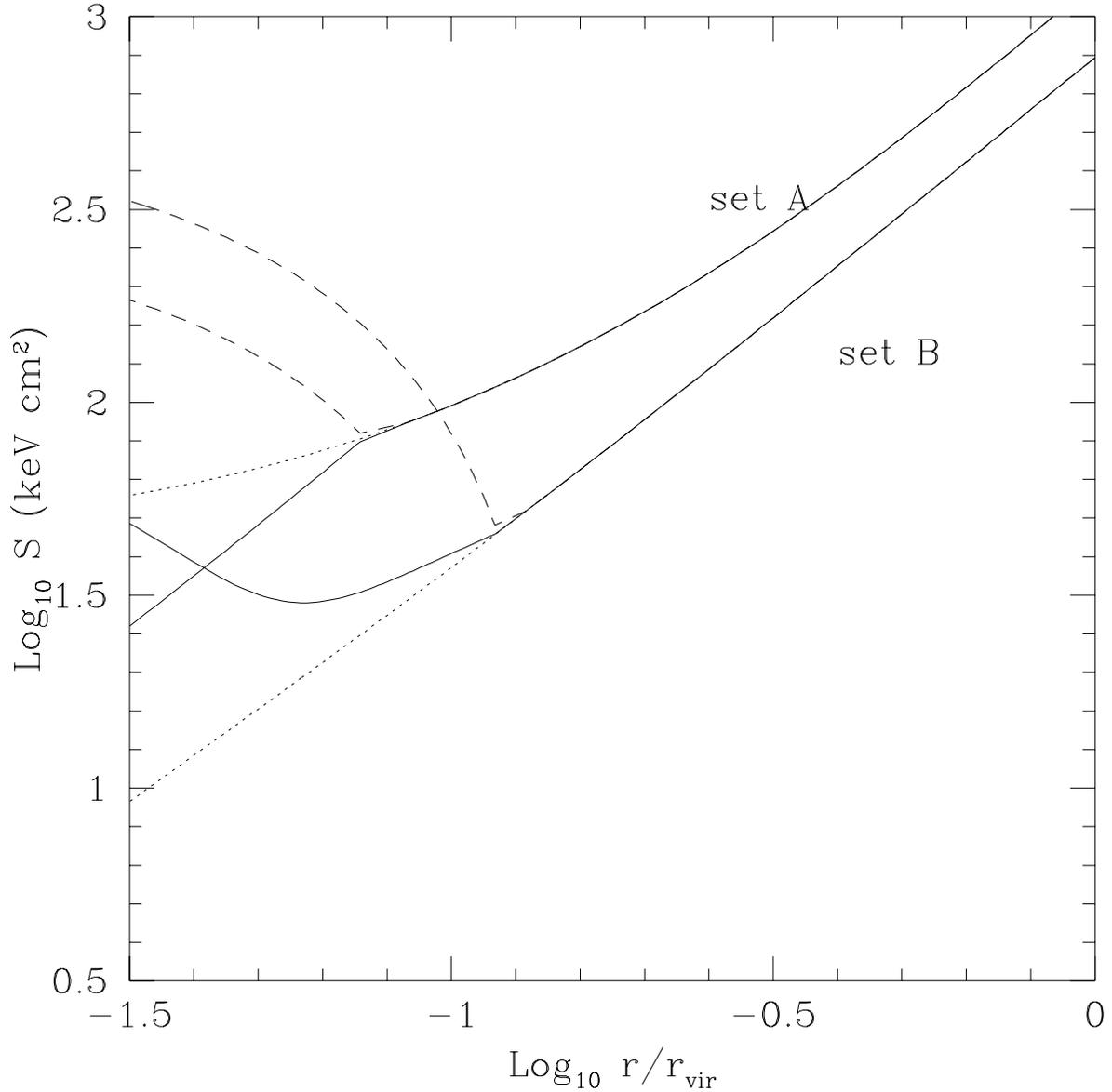}
}
{\vskip-3mm}
\caption{
Entropy profiles for the cases corresponding to those in Figure 2
are shown. The upper set of curves are for set A and the lower set
for set B default profiles. Dotted, Solid and long dashed lines show the
entropy for the default profile, the profile for the eigensolution
and the profile for the highest mass flux rate respectively.
}
\end{figure}

This difference is  also manifested in the mass flux rates. Figure 4 shows
the mass flux rates for the cooling eigensolutions (lowest entropy cases),
with solid line for set A and dotted line for set B. Observational data
points from White \etal (1997) 
are also shown for comparison, although we note that in light of the new
measurements (see B\" ohringer \etal 2002) these data points may be overestimates
for the mass flux. The mass flux is plotted against
the emission weighted temperature (see below). As expected from our definition
of `cooling eigensolutions', these solutions should provide the lowest mass
flux rate for observed cooling flows, since flows with mass flux lower than
this would correspond to holes in X-ray. The locus of these solutions for set A
indeed seems to provide a lower envelope for the data points. The mass flux
rate for the lowest entropy cases for set B profiles are larger than these.

Since the appropriate mass flux rate for our highest entropy flows are 
$\sim 100$ times larger than those for the cooling eigensolutions, from the
curves in Figure 4, we find that these highest entropy flows are characterized
by mass fluxes in excess of $1000$ M$_{\odot}$ per yr, for clusters with
emission weighted temperature $T_w \le 1$ keV.

Going back to the entropy profiles in Figure 3, the long dashed lines show the
highest entropy cases.
It is seen from this figure that there is a variety of entropy profiles for the
gas undergoing cooling flow depending on the boundary conditions. We wish to
point out that all these profiles have only one free parameter, once the
default profile is given, and that is the age of the system (assumed to be
$10^{10}$ and $1.5 \times 10^{10}$ yr here). 
In other words, there is no single entropy threshold
given the time allowed for the gas to cool, but that there is a large number
of possibilities spanning a large range in entropy, depending on the mass flux.

\begin{figure}
\centerline{
%{\vskip-4mm}
\epsfxsize=1.0\textwidth
\epsfbox{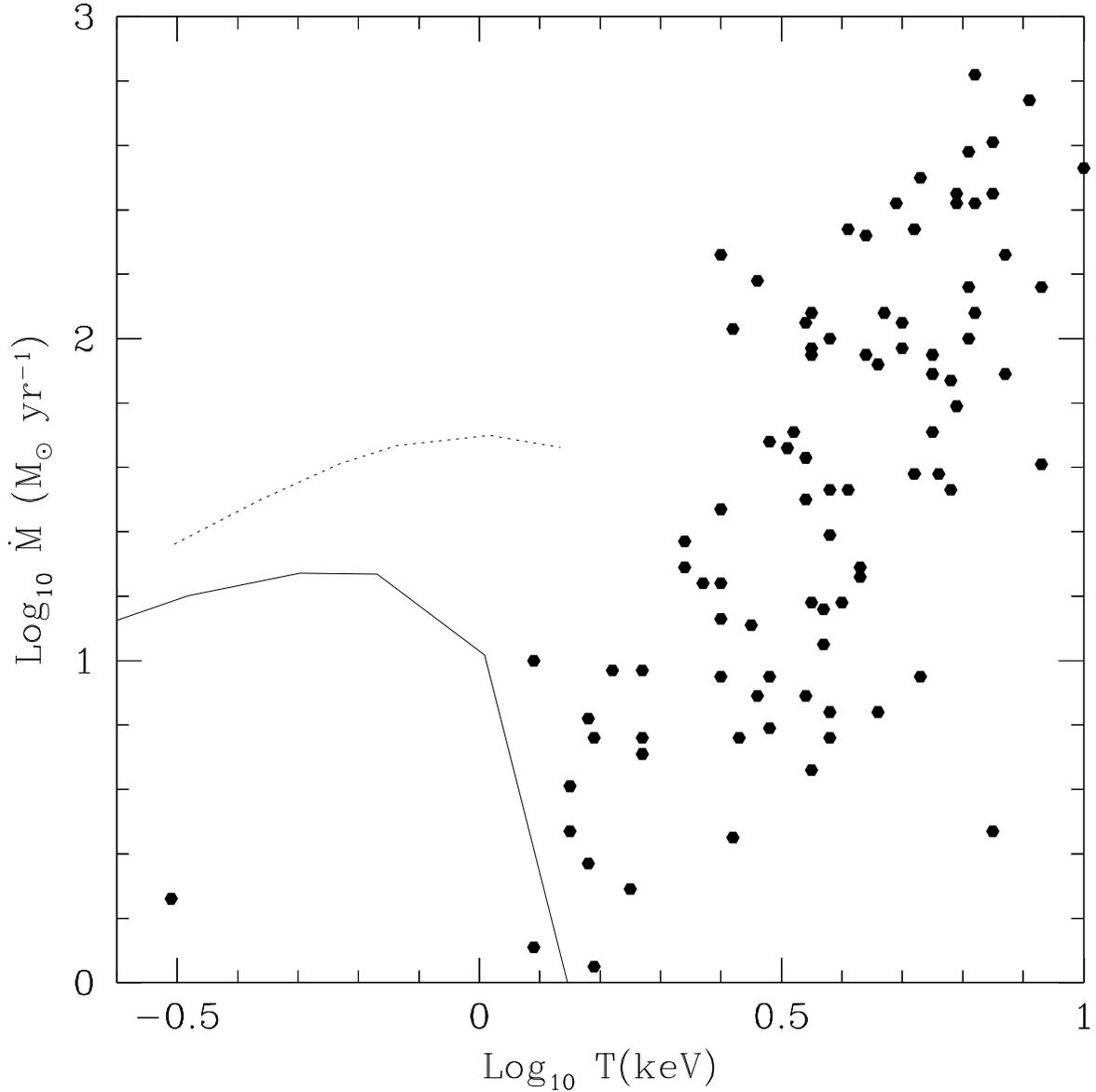}
}
{\vskip-3mm}
\caption{
Mass flux rates are plotted against (emission weighted) cluster temperature
for the eigensolutions for set A (solid line) and set B default profiles
(dotted line), for a cooling time of $10^{10}$ yr. The mass fluxes for a 
cooling time of $1.5 \times 10^{10}$ yr are somewhat larger than those plotted
here. Data points are from White \etal (1997).
}
\end{figure}

\subsection{X-ray luminosity}
We compute the bolometric X-ray 
luminosity, and emission weighted temperature 
corresponding to the profiles discussed above (in the band $0.5\hbox{--}10$ keV),
%in the $0.5\hbox{--}2.0$ keV band, 
using the Raymond Smith code, for
a metallicity of $Z/Z_{\odot}=0.3$. We compute the luminosities within the fiducial
radius $r_{200}$ as mentioned earlier to facilitate better comparison with data,
as previous authors have done. The X-ray luminosity and emission weighted
temperature is computed as,
\bea
L_x &&=\int_0^{r_{200}} 4 \pi r^2 n_i(r) n_e(r) \epsilon_{0.5-10} dr \, \nonumber\\
T_w &&={\int_0^{r_{200}} 4 \pi r^2 n_i(r)n_e(r) \epsilon_{0.5-10} T(r) dr \over
\int_0^{r_{200}} 4 \pi r^2 n_i(r)n_e(r) \epsilon_{0.5-10} dr} \,,
\eea
where $n_i,n_e$ represent the ion and electron density and $\epsilon_{0.5-10}$
denotes the emissivity relevant for the $0.5\hbox{--}10$ keV band.
We present the results in Figure 5 where the top panel shows the result for
set A and the bottom panel shows the result for set B default profiles.
The left panels show the results for a cooling time of $10^{10}$ yr and the right
panels for a cooling time of $1.5 \times 10^{10}$ yr.
In the figure, the dotted lines again show the luminosities for the default profiles,
the solid lines show the case for the eigensolutions (lowest entropy) and
the long dashed lines show the case for very large mass flux rates. 
Dotted lines between the solid and dashed lines connect the cases of equal
cluster masses. Data points
from Helsdon \& Ponman (2000) (filled circles) and Arnaud \& Evrard (1999)
(empty circles), 
corrected for our choice of
$h=0.65$ are also shown for comparison.
The luminosities plotted here from our calculations
 are not `cooling flow' corrected, since
the data points  are also not corrected for
any possible cooling flows.

\begin{figure}
\centerline{
%{\vskip-4mm}
\epsfxsize=1.0\textwidth
\epsfbox{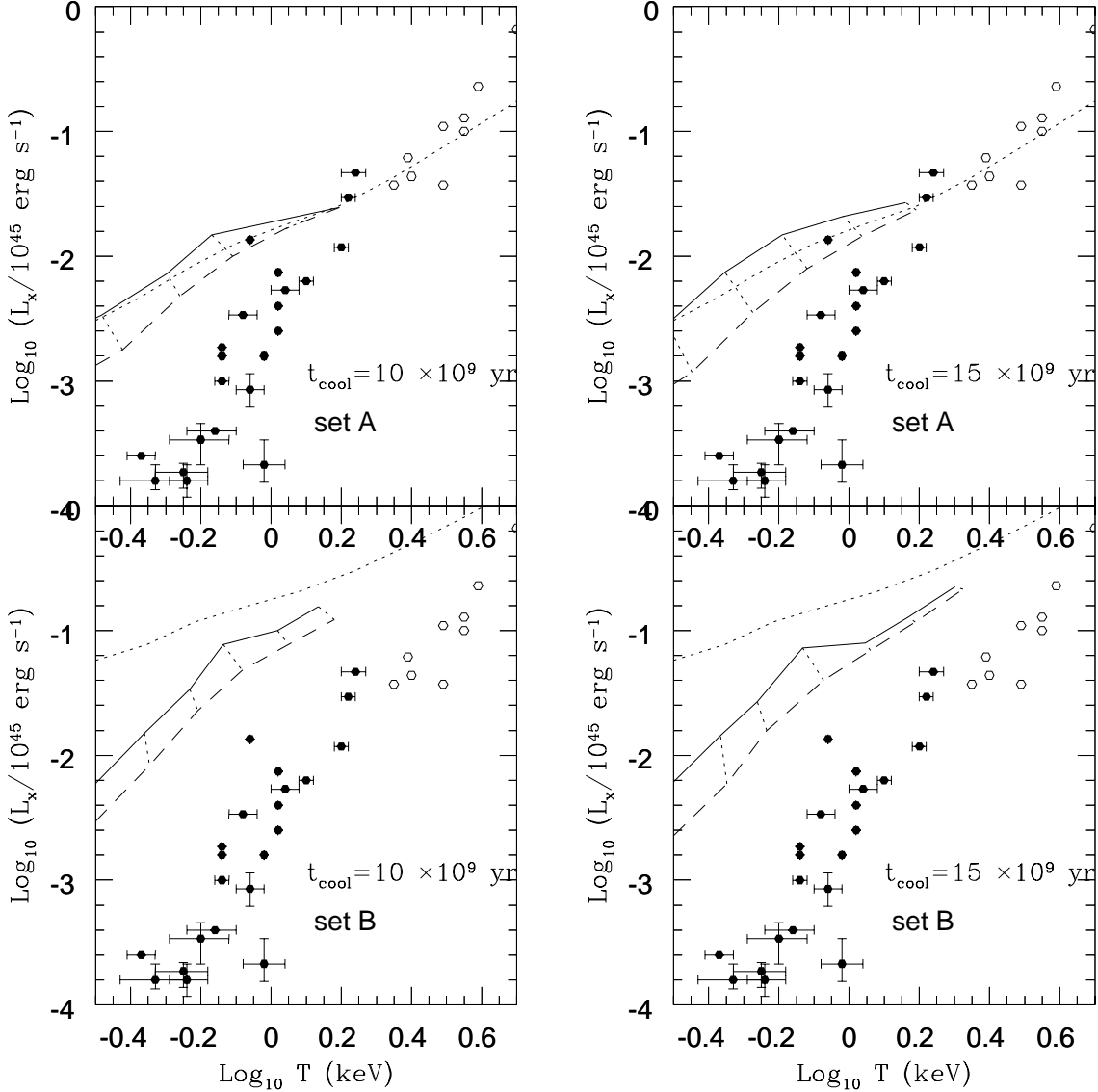}
}
{\vskip-3mm}
\caption{
X-ray luminosities are plotted against the emission weighted cluster
temperatures for set A (top panel) and set B (bottom panel) default
profiles, for cooling time of $10^{10}$ yr (left panels) and $1.5 \times
10^{10}$ yr (right panels).
Solid, dotted and long dashed lines have the same meaning as in
previous figures. Dotted lines between the solid and dashed lines 
connnect the cases of equal cluster masses.
Data points are from Helsdon \& Ponman (2000) and
Arnaud \& Evrard (1999)
corrected for our choice of $h=0.65$.)
}
\end{figure}

\subsection{Entropy at $r=0.1 r_{200}$}

We next present the values of the entropy at the fiducial radius $r=0.1 \, r_{200}$
in Figure 6, with results for set A in the top panel and those for set B in
the bottom panel. The dotted lines in each case show the entropy for the default
profile, the solid line show the case for 
the eigensolutions and the long dashed lines show the
case of the highest entropy profiles. 
%In addition, we show by the short dashed
%line the entropy values at the cooling radius. 
The dot-and-dash line shows the
result of N-body simulation from Ponman, Cannon \& Navarro (1999), showing the
self-similar case, and the data points are also taken from this work.

Below this temperature, it is possible to raise (or even lower) 
the entropy level of the gas
undergoing cooling flow, depending on the mass flux, 
%to compare the theoretical prediction with the data given the large uncertainty in
%observations, although 
and it is possible that cooling flows in very poor clusters
may be responsible for raising the entropy of the gas to some extent.

\begin{figure}
\centerline{
%{\vskip-4mm}
\epsfxsize=1.0\textwidth
\epsfbox{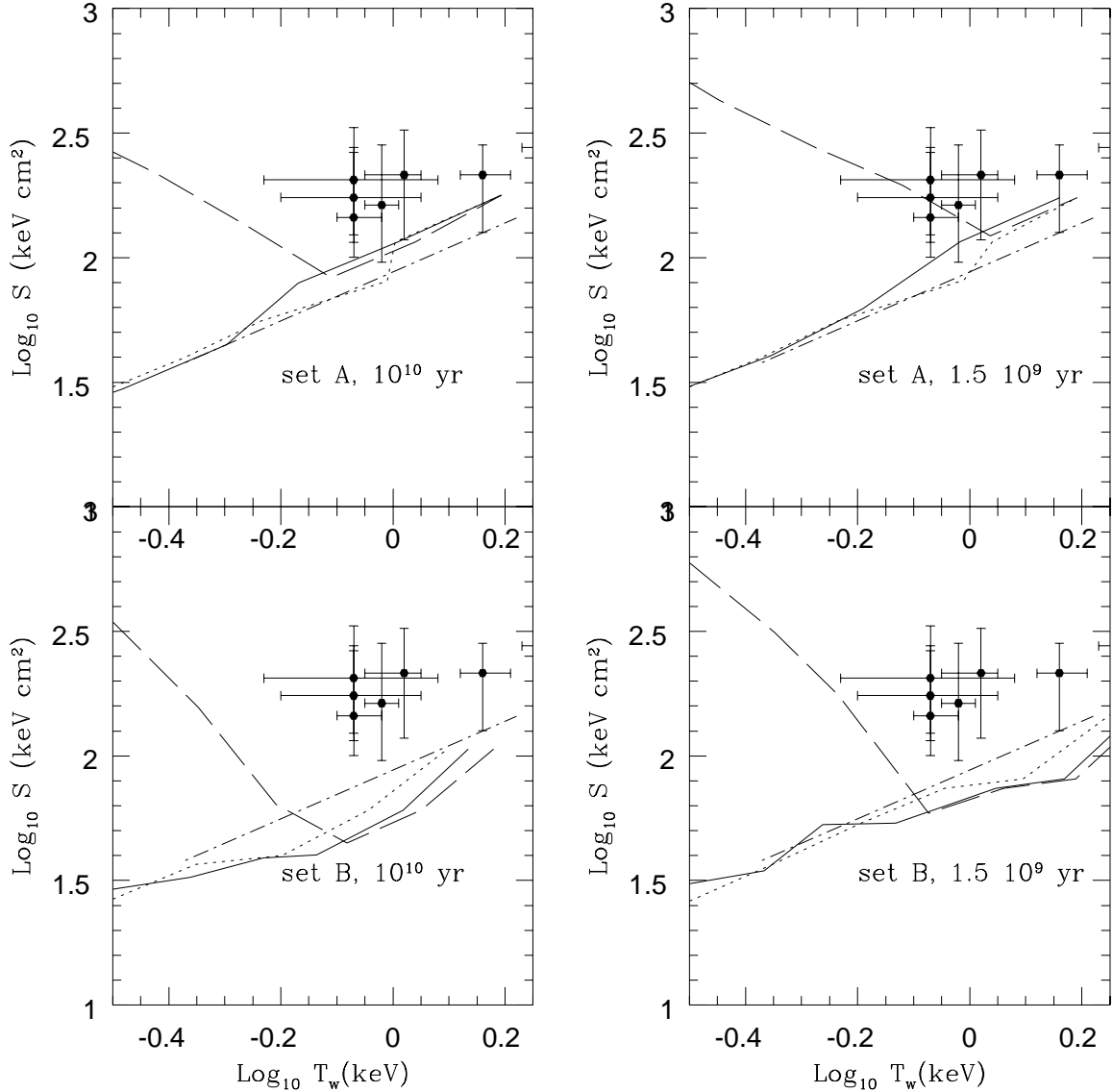}
}
{\vskip-3mm}
\caption{
Entropy at $r=0.1 \,r_{200}$ is plotted against emission weighted
cluster temperature for set A (top panel) and set B (bottom panel)
default profiles, for a cooling time of $10^{10}$ yr (left panels) and
$1.5 \times 10^{10}$ yr (right panels).
 Solid, dotted and long dashed lines have the same
meaning as in previous figures. 
%The short dashed line plots the
%value of the entropy at the cooling radius. 
The dot-and-dashed line
shows the results of the numerical simulation showing the self-similar
expectations, from Ponman \etal (1997).
}
\end{figure}

\section{Discussion}
We will discuss some of the implications of our results as well as a few 
uncertainties involved in our calculations in this section.

The results presented in the previous section show some of the difficulties of
explaining the observations with radiative cooling alone. To recapitulate
the premises of our calculations, we assumed that radiative
cooling must give rise to a gas flow towards the centre. Here we have
only studied steady and subsonic flows for simplicity, and motivated by the
fact that recent X-ray observations do not detect flows with very high mass
flux which require supersonic flows (except in the very vicinity of the centre)
and also flows with mass deposition (which would make it inhomogeneous) are not
detected. Let us discuss the implications of our results given these assumptions.

\subsection{Entropy}
Curves in Figure 6 show that  the default profiles do not differ much from
the self-similar case, although the set A profiles generally have larger entropy.
The region bounded by the solid and long dashed lines show the possible values
of entropy at $r=0.1 \, r_{200}$ for cooling flows with different mass fluxes,
with flows of larger mass flux rate occupying regions with larger entropy.
As the curves show, these solid and dashed lines merge with the  dotted line
of the default profile at high temperatures. This is because of the simple
reason that for high temperature clusters, the cooling radius is smaller than
the fiducial radius $r=0.1 \, r_{200}$, and so cooling flows do not affect the
entropy measurement at this fiducial radius for high temperature clusters.

Since the entropy measurements are made at a given fiducial radius of
$r=0.1 \, r_{200}$, for
cooling flows to have any effect on the raising of the entropy, the cooling
radius {\it must} exceed this fiducial radius. We find that for both of
our choice of default profiles, the cooling radius is inside this fiducial
radius for $T_w \ge 1$ keV, for a cooling time of $10^{10}$ yr. 
For a longer cooling time of $1.5 \times 10^{10}$ yr, the set A profiles
provide a larger cooling radius, and can potentially explain the entropy measurements
given the uncertainties. 
%Of course, for very larg temperature clusters,
%the self-similar case is enough to explain the data. This however makes it 
%difficult to explain the data around $T_w \sim 1$ keV, where the self-similar
%expectations fall short of the observed data (see Figure 6).

Below this temperature, it is possible that cooling flows can enhance the
entropy of the gas given a large enough mass flux rate (shown by the long dashed
lines in Figure 6, corresponding to mass flux rates of $\sim 1000$ M$_{\odot}$
per yr). It is interesting to note that the maximum attainable entropy
(at $r=0.1 \, r_{200}$) is much larger than the entropy at the cooling radius
(as seen from the example of entropy profile in Figure 3). 
The entropy at the cooling radius is obtained
by setting the cooling time of the gas equal to the age of the system, as was
calculated by Voit \& Bryan (2002). Flow of gas under the influence of the
gravitational potential can heat up the gas substantially and raise the 
entropy inside the cooling radius, which is clearly shown in the example portrayed
in Figure 3.

\subsection{X-ray luminosity}
From the results in Figure 5, we find that
although cooling flows decrease the luminosities to some extent,
the calculated X-ray luminosities are still larger than
observed luminosities. The inclusion of cooling flows decreases 
the X-ray luminosities
to some extent for set A profiles, and to a large extent for set B profiles.
This effect steepens the $L_x$ vs. $T$ relation compared to those of the
default profiles. The
lowest attainable X-ray luminosities correspond to the long dashed lines, which
show the case of flows with $({t_{fo} \over t_{co}}) =0.01$, and with
 mass flux rate of $\ge 1000$ M$_{\odot}$ per yr. Increasing the cooling
time to $1.5 \times 10^{10}$ yr decreases the luminosities somewhat more, but
are still larger than the observed luminosities.
We therefore find that flows with mass flux rates in excess of $1000$ 
M$_{\odot}$ per yr
are required to explain the X-ray luminosities with radiative cooling alone.

Recently Bryan (2000) and Wu \& Xue (2002a) showed that one can explain the X-ray
observations by eliminating the gas in the central region (defined by the cooling
region by Wu \& Xue (2002b)) and replacing this gas by the outer gas, {\it keeping
the entropy constant}. In reality, however, the flow that would ensue as a result
of cooling flows will not keep the entropy of the gas constant, since it is not
an adiabatic process. It is however possible to neglect the cooling of the gas,
and therefore assume an isentropic transport of the gas, {\it if the flow time
is much smaller than the cooling time of the gas}. This situation, however,
corresponds to flows with very large mass flux rates, as shown by equation
(12) that defines the flow time. We therefore reach the conclusion that flows
with very large mass flux rates are needed to explain the observations.

\subsection{Limitations}
In reality the cooling radius is an increasing function of time, and the results
presented above pertain to a simplified treatment of steady cooling flows, with
the mass flux rate being constant in time and for all radii. One expects the
mass flux rates to vary in time, as shown by the study of self-similar cooling flows
by Bertschinger (1989), or the numerical simulation by Knight \& Ponman (1987).
Also, the cooling flows are expected to be disrupted from time to time by mergers
of subclusters which would change the mass flux rate, as seen in the numerical
simulation by Knight \& Ponman (1987). Admittedly our simplified treatment 
cannot capture
these aspects of cooling flows, but the conclusions reached in the previous
paragraphs are expected to be general and robust. 

Also, there has been a recent surge of interest
in the study of cooling flows in light of the recent X-ray observations, and
a number of authors have pointed out the intimate connection between active
galaxies and cooling flows (e.g, Churazov \etal  2001;
Binney \& Kaiser 2002, and references therein).
For example, Binney \& Kaiser (2002) pointed out that as the entropy is
continually lowered by cooling flows, it will trigger some nuclear activity
in the centre from time to time. They argued that this would make flows with
a large drop in temperature in the centre less probable for detection.
We did not include any heating source in our cooling flow equations so that
we could study the effect of only radiative cooling. In fact our results
suggest that additional heating sources, possibly from active galaxies,
would be needed to explain the observations. 

%For example,
%the fact that the cooling radius is shorter than the fiducial radius for
%entropy measurements ($0.1 \, r_{200}$) for $T_w \ge 1$ keV should not depend
%on the complications mentioned above, as well as the results for X-ray
%luminosities. 

\subsection{Uncertainties}
We wish to discuss some of the uncertainties pertaining to our 
calculations. First,
there is an uncertainty of the background dark matter density profile. 
In the preceding
sections, we have only discussed the case of the universal profile 
with the value of
$\alpha=1$. We have also calculated the lower and upper bounds of the 
X-ray luminosity
in the case of $\alpha=1.5$,  
and found that results differ only by a small amount.
Recently, however, Kelson \etal (2002) have pointed out that it is even
possible for the value of $\alpha$ to be much smaller than unity. Also, it was
shown by Lloyd-Davies \etal (2002) that the choice of the concentration parameter
can introduce some uncertainty in the calculations of X-ray properties of the gas.
Voit \etal (2002) also showed the choice of the concentration parameter can 
shift the results to some extent.
We based our calculations on two different sets of default profiles for these
uncertainties, to determine the effects of the assumption of initial profiles on 
the final result.
 Although the details of cooling flows for our two choices 
of default profiles
are somewhat different, the final conclusions do not differ much, suggesting
that our conclusions are robust.

\section{Summary}

We have studied cooling flow solutions for two different sets of default profiles,
one with gas obeying a polytropic equation of state (set A)
and the other in which the gas
density is proportional to that of the dark matter (set B). 
We have simplified the
cooling flow equations to the case of steady, homogeneous and 
subsonic flows. Within the purview of
such flows, we find that :

(a) cooling flows can either increase or decrease the entropy of the gas compared
to the default (`no-cooling') profile depending on the
mass flux rate. In other words, cooling for a given duration does not introduce 
a given entropy threshold for the gas; 
the entropy profile depends on the mass flux rate.

(b) For a cooling time of $10^{10}$ yr, the cooling radius
exceeds the fiducial radius $r=0.1 \, r_{200}$ only for clusters
with $T_w \le 1$ keV, 
which makes it difficult for cooling flows to have any effect
on the entropy at this radius for clusters with $T_w$ of order of a few
keV, where the expected entropy from the self-similar case falls short of the
observed values. Increasing the cooling time to $1.5 \times 10^{10}$ yr can
increase the cooling radius for clusters beyond $T_w \sim 1$ keV, 
and can potentially
explain the observations.
We also find that cooling flows can increase the entropy at this fiducial radius
for lower temperature clusters.

(c) Cooling flows that are caused by radiative cooling of gas in the central
region can decrease the X-ray luminosities for poor clusters and steepen the
relation between $L_X$ and $T$ compared to the self-similar expectation, but
flows with mass flux rates in excess of $\sim 1000$ 
M$_{\odot}$ per yr are
required to explain the observations if heating sources are not 
taken into account.

\bigskip
\noindent
{\bf Acknowledgement}
I am indebted to  Drs. Xiang-Ping Wu and Yan-Jie Xue for their gracious
help with using the Raymond-Smith code. The comments of the anonymous referee
which helped to improve the paper are also acknowledged with thanks.

\section{Appendix A}

\begin{figure}
\centerline{
%{\vskip-4mm}
\epsfxsize=1.0\textwidth
\epsfbox{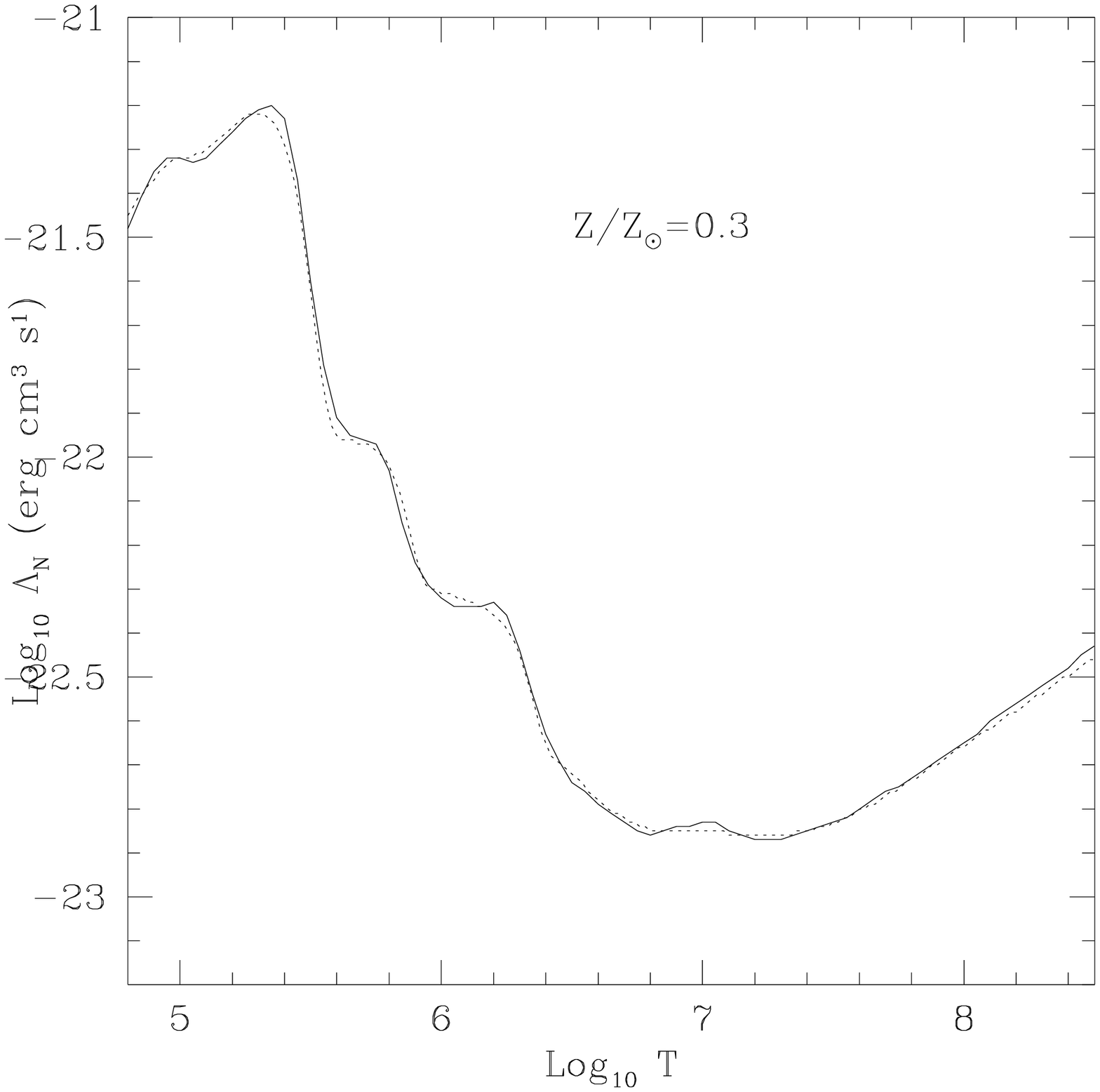}
}
{\vskip-3mm}
\caption{The cooling function of Sutherland \& Dopita (1993) is shown
for $Z/S_{\odot}=0.3$ with solid line, along with the analytical fit
described here with dotted line. 
}
\end{figure}

We have found
that the following function fits the Sutherland \& Dopita (1993)
cooling function for $Z/Z_{\odot}=0.3$ very accurately,
with an accuracy of $5\%$ in the range $4.8 < \log _{10}T <8.5$, which we
have used in our calculations,
\bea
{\Lambda_N \over 10^{-23} \, {\rm erg} \, {\rm cm}^3 \, {\rm s}^{-1}}
&=&0.02\, ({T \over 2 \times 10^4})^{4.8} \, 
(\exp -({T \over 8.3 \times 10^4})^4) \nonumber\\
&&+24.\,({T \over 8 \times 10^4}) \, 
(\exp -({T \over 2.75 \times 10^5})^5) \nonumber\\
&&+5.5 \, (\exp -({T \over 7.4 \times 10^5})^{8.5}) \nonumber\\
&&+150. \,T ^{-0.3} \, (\exp -({T \over 2.2 \times 10^6})^8) \nonumber\\
&&+2.1 \, (\exp -({T \over 3. \times 10^6})^{1.8}) \nonumber\\
&&+6.\times 10^{-7} \,T \, (\exp -({T \over 2.8 \times 10^6})^7) \nonumber\\
&&+0.35 \, ({T \over 10^6})^{0.4} \,,
\eea
We plot this function along with the net cooling function from Sutherland \&
Dopita(1993)
for $Z/Z_{\odot}=0.3$, in Figure A.

\end{document}